\documentclass[fleqn,10pt]{wlscirep}

\usepackage[utf8]{inputenc}
\usepackage[T1]{fontenc}
\usepackage{trace}
\usepackage{color}
\usepackage{amsmath}

\title{Generative Deep Learning Model for a Multi-level Nano-Optic Broadband Power Splitter}

\author[1,2]{Yingheng Tang}
\author[1,*]{Keisuke Kojima}
\author[1]{Toshiaki Koike-Akino}
\author[1]{Ye Wang}
\author[1]{Pengxiang Wu}
\author[1]{Mohammad Tahersima}
\author[1]{Devesh K. Jha}
\author[1]{Kieran Parsons}
\author[2]{Minghao Qi}

\affil[1]{Mitsubishi Electric Research Laboratories, 201 Broadway, Cambridge, MA 02139, USA}
\affil[2]{School of Electrical and Computer Engineering and Birck Nanotechnology Center, Purdue University, West Lafayette, IN 47907, USA }
\affil[*]{kojima@merl.com}


\begin{abstract}
We propose a novel Conditional Variational Autoencoder (CVAE) model, enhanced with adversarial censoring and active learning, for the generation of $550$~nm broad bandwidth ($1250$~nm to $1800$~nm) power splitters with arbitrary splitting ratio. The device footprint is $2.25 \times 2.25~\mu \textrm{m}^2$ with a $20\times20$ etched hole combination. It is the first demonstration to apply the CVAE model and the adversarial censoring for the photonics problems. We confirm that the optimized device has an overall performance close to $90\%$ across all bandwidths from $1250$~nm to $1800$~nm. To the best of our knowledge, this is the smallest broadband power splitter with arbitrary ratio.
\end{abstract}

\begin{document}
\flushbottom
\maketitle
\thispagestyle{empty}

\section*{Introduction}

Inverse design of photonic devices has been widely studied over the past years. Different optimization methods (direct binary search (DBS)~\cite{mrowca2018flexible,spelke1992origins,tenenbaum2011grow}, genetic algorithm~\cite{mccloskey1980curvilinear,smith2013sources}, particle swarm optimization~\cite{spelke1990principles}, etc.) have been used for different applications and have been shown to have unique strengths.
However, all of these methods always require a large amount of time and computing resources.
For example, a DBS optimization setup with $20 \times 20$ hole vectors has an extremely large number of possible combinations $(2^{400})$, which require many electro-magnetic (EM) simulations like finite-difference time domain (FDTD) simulations.
Machine learning, on the other hand, can learn complex input-output relationships by creating large models which can be then trained using large amounts of data. Taking inspiration from these algorithms, machine learning (including deep learning) is becoming more popular to assist the design process of photonic devices to accelerate the underlying optimization process in the past few years.
Recent success of deep learning in modeling complex input-output relationship in spatial-temporal data, has inspired the idea of intuitive physics engines that can learn physical dynamics in mechanics~\cite{chang2016compositional,lerer2016learning,battaglia2016interaction}, material discovery~\cite{sendek2018machine,gomez2018automatic,ghaboussi1991knowledge}, particle physics~\cite{radovic2018machine}, and optics~\cite{tahersima2019deep,liu2018training,ma2018deep,malkiel2018deep,peurifoy2018nanophotonic,sun2018efficient,liu2018generative,asano2018optimization,hammond2019designing,ma2019probabilistic,kojima2017acceleration,teng2018broadband}. 

Deep neural networks do require a large data pool to train and can be notoriously difficult to train for some system. However, once the network is trained properly, the approximate response would be almost instantaneous and hopefully accurate. A critical criterion for the application of such intuitive engines for physical systems is their generalization capability to problems beyond the training data set. Such generalization capability would enable use of neural network models as a forward design optimization engine that is trained on limited partially-optimized data set. 

In the silicon photonics field, there have been several attempts at combining the machine learning algorithms with the DBS design process. We developed an artificial intelligence integrated optimization process using neural networks (NN) that can accelerate optimization by reducing the required number of numerical simulations~\cite{kojima2017acceleration, teng2018broadband}. Also, Tahersima et al.~\cite{tahersima2019deep} used DNN in the inverse direction, i.e., use target performance data (such as transmission spectra) as input, and device design as output. However, the DNN network structure we used (i.e., ResNet) was one-to-one deterministic mapping, which generates only one certain device for every performance set. Another limiting factor of our previous demonstrations is that device consists of binary pixels (i.e., etch hole is present or not). To overcome the limitation of binary holes, we propose a multilevel pixel structure (i.e., multi etch hole dimensions), which is a more complex optimization problem and requires more sophisticated optimization algorithms. 

In the area of metamaterials, a few works have proposed to use a generative network for the pattern generation. A generative network generates a series of nearly optimum patterns based on random numbers. Liu et al.~has applied the Generative Adversarial Networks (GAN)~\cite{liu2018training}. and Ma et al.~has employed the Variational Autoencoder (VAE)~\cite{ma2019probabilistic} for their applications. Inspired by these works, we propose to utilize Conditional Variational Autoencoder (CVAE) in our power splitter application. One advantage that the VAE has comparing to the other network is that it models the probability distribution of the existing data so that it could generate new data from that distribution. 

By using the VAE, we can model the distribution of the splitters with different splitting ratios, and thereby allows generating novel patterns subject to this same distribution through data sampling. When coupled with conditions, VAE evolves into CVAE, and enables to produce patterns satisfying the given conditions. In our application, we use different hole sizes to express the appearances. In this way, the generated patterns can work better in the light guidance and make the generated devices more stable. Further, we add an adversarial block to isolate the pattern from the performance during the training, to further improve the performance. Our device footprints are $2.25 \times 2.25 \mu \textrm{m}^2$ with a $20\times20$ etched hole combination. We confirm that the optimized device has an overall performance close to $90\%$ across all the bandwidth from C-band to O-band (1250~nm to 1800~nm). To the best of authors’ knowledge, this is the smallest broadband power splitter with arbitrary ratio and  it is the first demonstration to apply the CVAE model for assisting the silicon photonics device design. We provide the following insight based on our numerical analysis through EM simulation software.

\section*{The device structures}

\begin{figure}[!h]
\centering
\includegraphics[width=14cm]{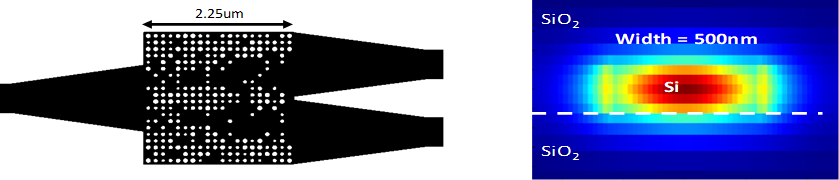}
\caption{Power splitter footprint $\&$ cross-section of the input/output waveguide.}
\label{fig:Device}
\end{figure}

Our device is a multi-mode interference (MMI) based power splitter with a footprint of $2.25 \times 2.25~\mu \textrm{m}^2$ with oxide cladding, a waveguide width of 500~nm and the height of 220~nm. We added a $20\times20$ Hole Vector (HV) to express the nanostructured hole configuration. The hole spacing is 112~nm, and the minimum and maximum hole diameters are 72~nm, and 40~nm respectively. The HV training data only consist of binary numbers initially and is obtained through direct binary search method. Note that the VAE models the probabilistic distribution for the different HV, each generated value of the HV is a Bernoulli’s distribution. In order to best reflect the result, different hole sizes are used to represent the probability of the appearance of etched holes at certain locations. Figure~\ref{fig:Device} shows the sample footprint of the power splitter.
Learning complex physical dynamics still remains a challenging problem and most often requires extensive training of the network.

\section*{The CVAE model}

\begin{figure}[!h]
\centering
\includegraphics[trim=0mm 0mm 0mm 3mm,clip,width=1\linewidth]{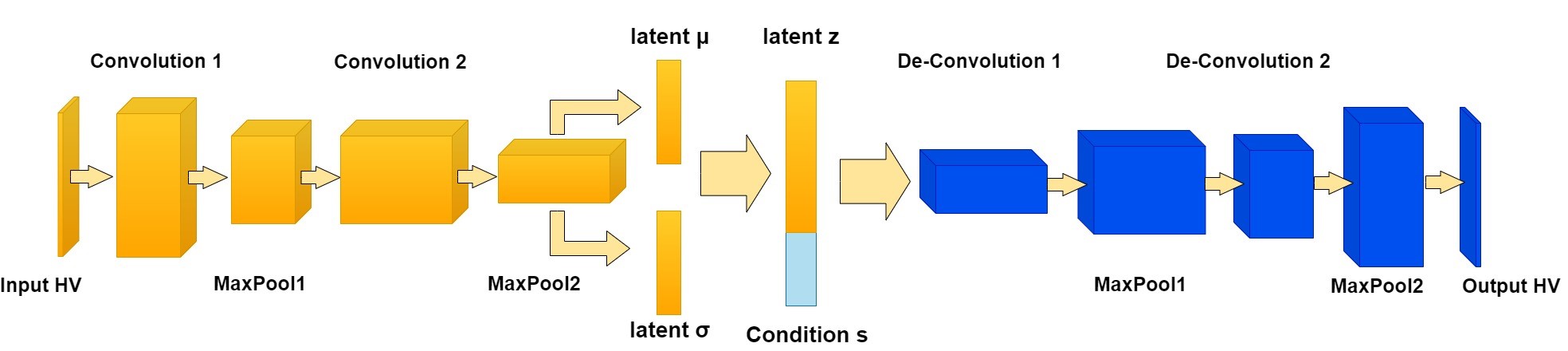}
\caption{The CVAE model structure. Here the input is the $20\times20$ hole vector. The first convolutional layer has $16$~channels with $3$~Kernels. The second convolutional layer has $32$~channels with $3$~Kernels. The condition s is the transmission and reflection spectra from $1.5~\mu \textrm{m}^2 -1.6~\mu \textrm{m}^2$ bandwidth, which is obtained from the FDTD simulation to form a $3\times21$ matrix.}
\label{fig:CVAE}
\end{figure}

We apply the variational auto-encoder~\cite{kingma2014adam} to extract the original $20 \times 20$ HV to certain types latent variables. The generative process of the VAE is as follows: a set of latent variable z (60 here in our application) is generated from the prior distribution. $P_\theta$ and the data x is generated by the generative distribution $P_\theta(x|z)$ with the condition on $z: z ~ P_\theta(z), x ~ P_\theta(x|z)$.

As shown in Fig.~\ref{fig:CVAE}, the original HV passes two convolutional layers and reduces to two sets of intermediate parameters of a Probability density function (Pdf), which are the mean $(\mu)$ and covariance $(\sigma)$. In order to make the back propagation possible for the network, the reparametrize trick is applied, which is shown in the following equation:
\begin{equation}
\begin{gathered}
z^i = \mu^i + \sigma^i \times \varepsilon
\end{gathered}
\end{equation}
Here $\varepsilon$ is a standard Gaussian distribution. Then reparametrized latent variable $z$ is concatenated with the condition parameter $s$ to deconvolute back to the HV.
The loss function that is used here are constructed by two parts: The Cross-Entropy loss between the original HV $x$ and the decoded HV $y$, and the Kullback–Leibler (KL) divergence between the encoder and the decoder. The equation of loss function is shown as the following: 
\begin{equation}
\begin{aligned}
Loss =  &-[y\log x+(1-y)\log (1-x)]\\
	  &+\frac{1}{2} \sum\limits_{j=1}^{J}[1 + \log (\sigma ^2_{zj}) - \mu ^2_{zj} - \sigma^2_{zj}]
\end{aligned}
\end{equation}

\begin{figure}[!ht]
\centering
\includegraphics[width=10cm]{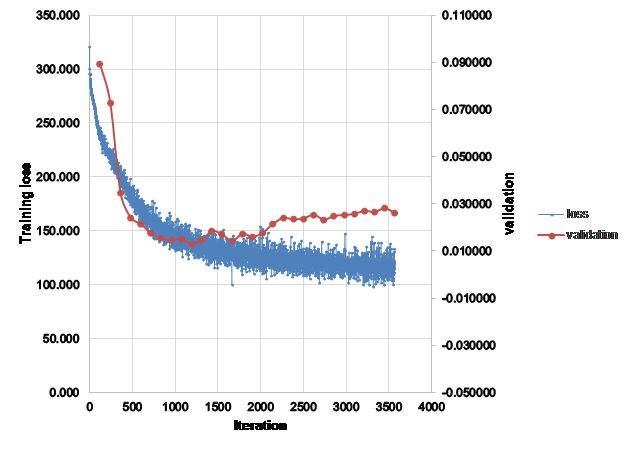}
\caption{The training loss and validation vs. iteration}
\label{fig:training}
\end{figure}

A convolutional neural network (CNN) is shown to be effective in handling the geometrical input data~\cite{tahersima2019nanostructured,liu2018generative,ma2019probabilistic,asano2018optimization}. Here we use two convolutional layers and the channel for them are $16$ and $32$, and the max pulling is $2$, after that there is one fully connected layer to reduce the latent variable to $63$. Then concatenate with the performance data and feed them into the decoder. Figure~\ref{fig:training} shows the training loss and the validation verse the training iterations. The solid line is the training loss and the dots are the trend for validation. The validation is calculated by using the EM simulation to verify the Figure of Merit of the generated pattern. The figure of merit is calculated as following: 

\begin{equation}
\begin{aligned}
\mathsf{FOM} = 1 - 10 \times \Bigl[\int_{a}^{b} |T_1(\lambda) - T_{1}^\star(\lambda)|^2 d\lambda + \int_{a}^{b} |T_2(\lambda) - T_{2}^\star(\lambda)|^2 d\lambda+ \int_{a}^{b} \alpha \times R^2(\lambda) d\lambda \Bigr],
\end{aligned}
\end{equation}
where $T_1(\lambda)$, $T_2(\lambda)$, $R(\lambda)$, and $[\cdot]^\star$ denote transmissions of output ports 1 and 2,  reflection at input port at a given wavelength $\lambda$, and corresponding target values, respectively. $\alpha$ is the weighting factor where $\alpha = 4$ is used to balance between the contributions from the transmission and the reflection. We take the average of $FOM$ over the FDTD simulation spectral range. The coefficient alpha used here is $4$ to make the reflection comparable to the transmission. For an ideal power splitter, the $FOM$ should be $1$. The plot shows that the training result is optimal when the iteration is around $1200$.

\section*{Experimental Results}
\begin{figure}[ht]
\vspace{-5pt}
\centering
\includegraphics[trim=0mm 5mm 0mm 4mm,clip,width=0.9\linewidth]{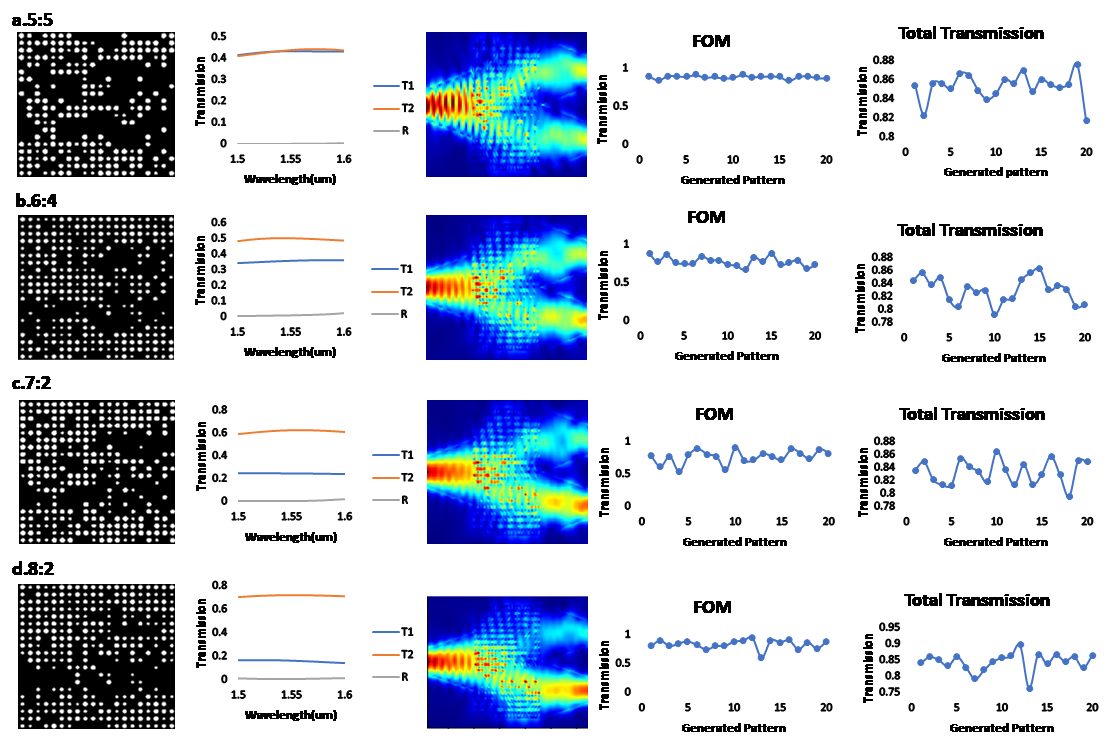}
\caption{The deivce simulation generated by the CVAE model over the 100nm bandwith (1500nm-1600nm)}
\label{fig:SimCVAE}
\end{figure}

After the training, we have tested the generator by using it to generate different devices. Since the VAE analyze the probabilistic distribution for the different HV, which makes each generated value of the HV a Bernoulli’s distribution. In order to best reflect the result, different hole sizes are used to represent the probability of the appearance of etched holes at certain locations. In order to verify the effectiveness of the generator, we choose 4 different devices with the different splitting ratios $(5:5, 6:4, 7:3, 8:2)$, Fig.\ref{fig:SimCVAE} shows the results that generated by the model and the FDTD verification for those. Here we use two metrics to represent the performance of the generated devices. First is the Figure of Merit (FOM), which can be expressed with the equation from the last section. The second is the total transmission. The third one is the Hamming distance for the generated devices. The left three columns are the hole pattern, the transmission plot and the beam propagation plot for one randomly generated result. 
The right two columns are the plots for the Figure of Merit and the total transmission for 20 randomly generated devices. The average $FOM$ is around 0.02 and the total transmission is at around $85\%$ across the the spectrum between 1500nm and 1600nm, which proves the generation capability of the model.

\begin{figure}[!ht]
\centering
\includegraphics[trim=0mm 0mm 0mm 3mm,clip,width=1\linewidth]{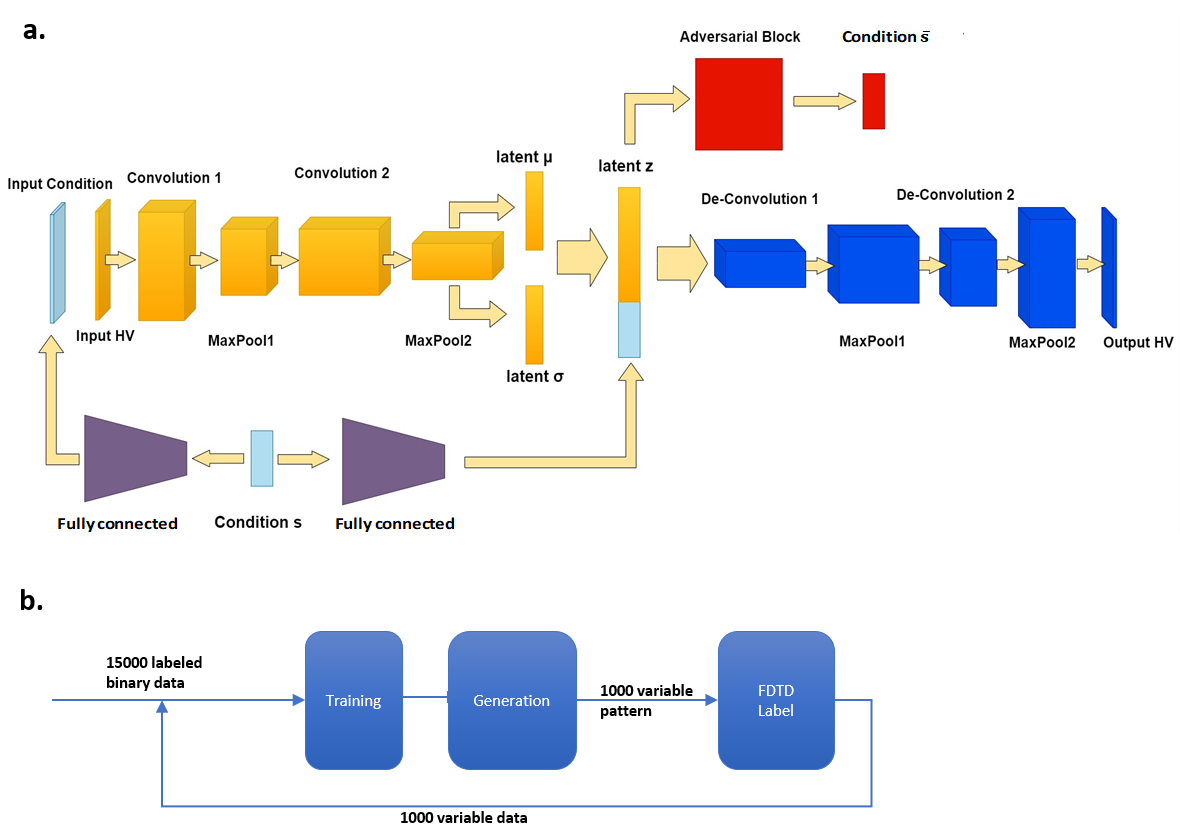}
\caption{(a) The updated CVAE model with adversarial block, and (b) the schematic of active learning.}
\label{fig:A-CVAE}
\end{figure}

\section*{The adversarial block and the active learning process}
The mechanism of applying the trained CVAE model is to feed the trained decoder with the desired condition along with the latent variable following the normal distribution. Ideally, the latent variables should obey the normal distribution $N(0,1)$. However, when the input pattern information gets encoded during of training of the preliminary model, some performance data may also be encoded into the latent variable as well. This will cause the distortion of the latent variables extracted and may cause degradation of the device performance for the generated pattern. To further improve the performance of the generator, the adversarial block was added. We add an adversarial block to isolate the latent variable z from the nuisance variations s (the performance data) in order to fit the device distribution better~\cite{lample2017fader, wang2019invariant, ozdenizci2019transfer}. Figure~\ref{fig:A-CVAE} shows the network structure of the adversarial CVAE. we use a decoder structure to expand the performance feature s into a $20\times20$ matrix and then combine with the original $20\times20$ hole vector to form a $2$-channel input, then process it through two convolution layers. One additional step is when the latent z variable is extracted, it will also be fed into an adversarial block to generate $\bar{s}$. The updated loss function is shown as following:

\begin{equation}
\begin{aligned}
Loss =  &- [y_n\log x_n+(1-y_n)\log (1-x_n)]\\
	  &+ \frac{1}{2} \sum\limits_{j=1}^{J}[1 + \log (\sigma ^2_{zj}) - \mu ^2_{zj} - \sigma^2_{zj}]\\
	  &- \beta \; MSE\_LOSS(s,\bar{s})
\end{aligned}
\end{equation}
Where
\begin{equation}
MSE\_LOSS = \frac{1}{n} \sum\limits_{i=1}^{n} (s_i-\bar{s}_i)^2
\end{equation}
The loss function has two parts. The first is the VAE loss which contains the Binary Cross-Entropy loss and the KL divergence. The second part is the MSE loss of the adversarial block. Since the condition information contained in the latent variable z needs to be minimized, the MSE loss between $s$ and $\bar{s}$ needs to be maximized. A complete update of the network needs to iterations. The first iteration is to backpropagate and update the CVAE model based on the loss function stated above. The second iteration is to update the adversarial block solely based on the MSE loss between $s$ and $\bar{s}$. During the training, we found when $\beta$ is 5, the generator gives the best performance. Figure 6 shows the latent variable distribution for 4 different types of devices.  The original latent variables are in dimension of 63 and the t-distributed Stochastic Neighbor Embedding (t-SNE) method is used to reduce the dimension to 2 for better visualization. The figure clearly shows that with adversarial censoring, all the latent variables obey the Normal distribution N(0,1), which is expected.\\

\begin{figure}[!ht]
\centering
\includegraphics[trim=0mm 0mm 0mm 3mm,clip,width=1\linewidth]{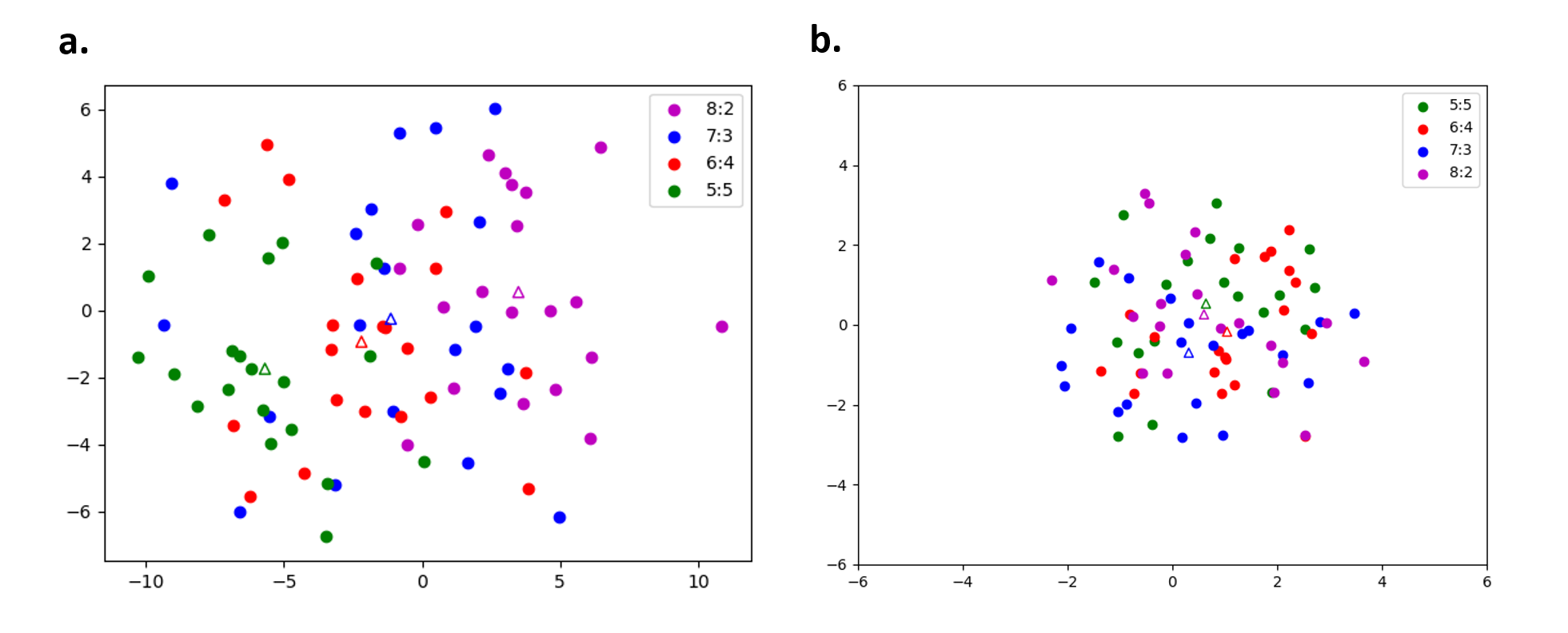}
\caption{t-SNE output of the latent variables. The latent variables are obtained from 4 different types of devices. We use the t-SNE to reduce the dimension to 2 for better visualization. The triangle markers are the centroid for each device group. 6a shows the latent variables for the CVAE model, which shows clear clustering. 6b shows the latent variables for the CVAE model, where all the latent variables obey the Normal distribution $N(0,1)$.}
\label{fig:Figure_t_SNE}
\end{figure}

The second method we are using to improve the model's performance is active learning. The process is showing in the Figure 5b. We first train a preliminary model using the original binary training data. Then, preliminary model is used to generate 1,000 variable hole size patterns with different splitting ratios. With finite-difference time-domain(FDTD), we label them to append into the training data for the second round.
There is a significant boost in terms of performance after the we apply the adversarial block along with the active learning. The simulation results are shown in the next section.

\begin{figure}[!ht]
\centering
\includegraphics[trim=0mm 0mm 0mm 3mm,clip,width=1\linewidth]{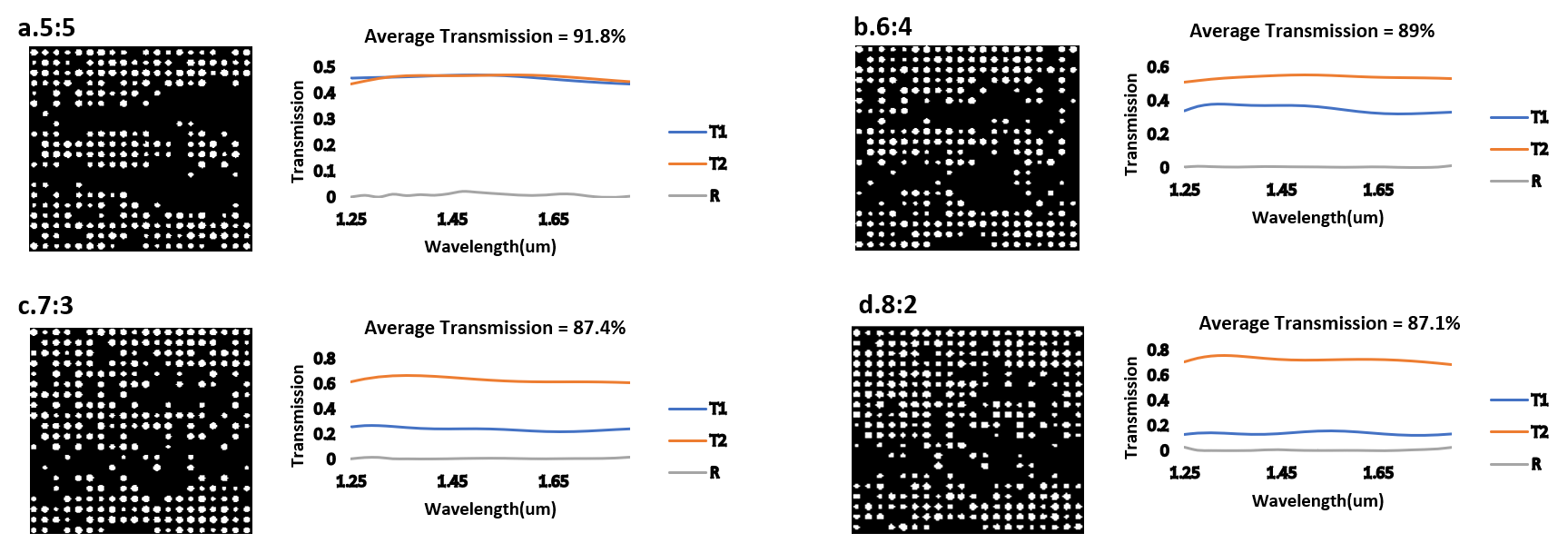}
\caption{FDTD results (transmission and reflection) of the generated patterns via the active learning assisted adversarial CVAE model.}
\label{fig:A-CVAE-Results}
\end{figure}

\begin{figure}[!h]
\centering
\includegraphics[trim=0mm 0mm 0mm 3mm,clip,width=1\linewidth]{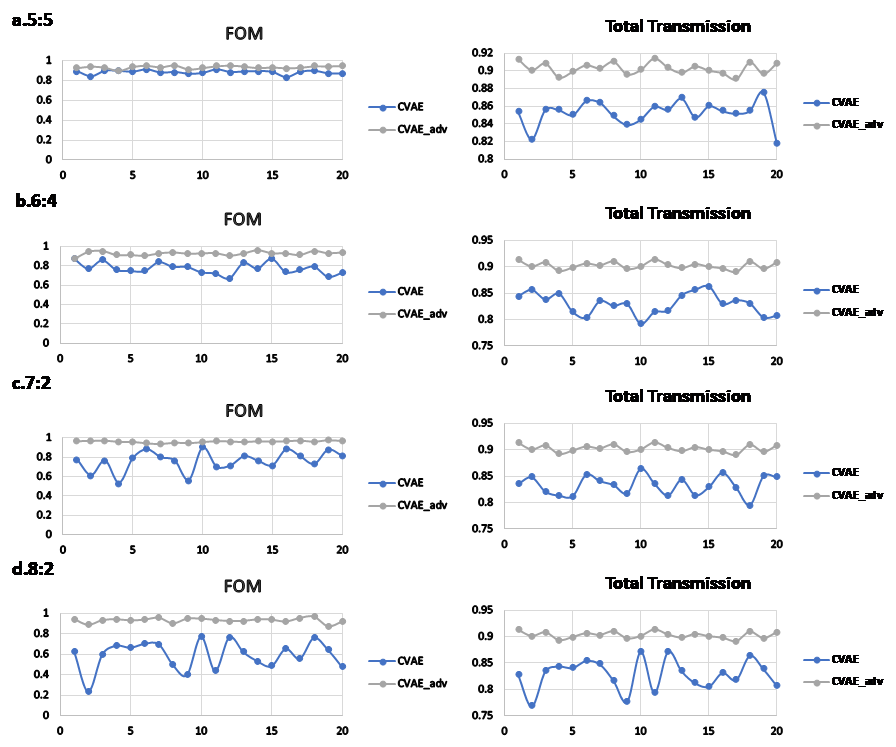}
\caption{Result of the generated pattern from the adversarial-CVAE model}
\label{fig:A-CVAE-pattern}
\end{figure}

\section*{Updated experiment result}

We use the updated generator to generate the same for types of devices (splitting ratio of $5:5$, $6:4$, $7:3$ and $8:2$) to make a comparison to the previous model. Same metrics are applied here to show the device performance which are the $FOM$ and the total transmission.
Figure~\ref{fig:A-CVAE-Results} shows the simulation result for the detailed transmission and reflection for the 4 types of devices that are generated by the ACVAE model. The reflection is smaller than -20dB and the achieved transmission is larger than $87\%$ across the bandwidth between $1250$~nm-$1800$~nm.
Figure~\ref{fig:A-CVAE-pattern} shows the results between the preliminary model and the updated model. Twenty random patterns are generated for each type of the device. The plot shows a boost of improvement from the previous model. For the 1:1 device, the total transmission increased from $85\%$ to $90\%$, and the FOM drops from $~0.015$ to $~0.005$. For the 6:4 devices, the total transmission increased from $83\%$ to $90\%$, for the ratio of $7:3$ and $8:2$, the total transmission has been increased from $85\%$ to ~$93\%$ across almost all randomly generated patterns. 

\begin{figure}[!h]
\centering
\includegraphics[trim=0mm 0mm 0mm 3mm,clip,width=1\linewidth]{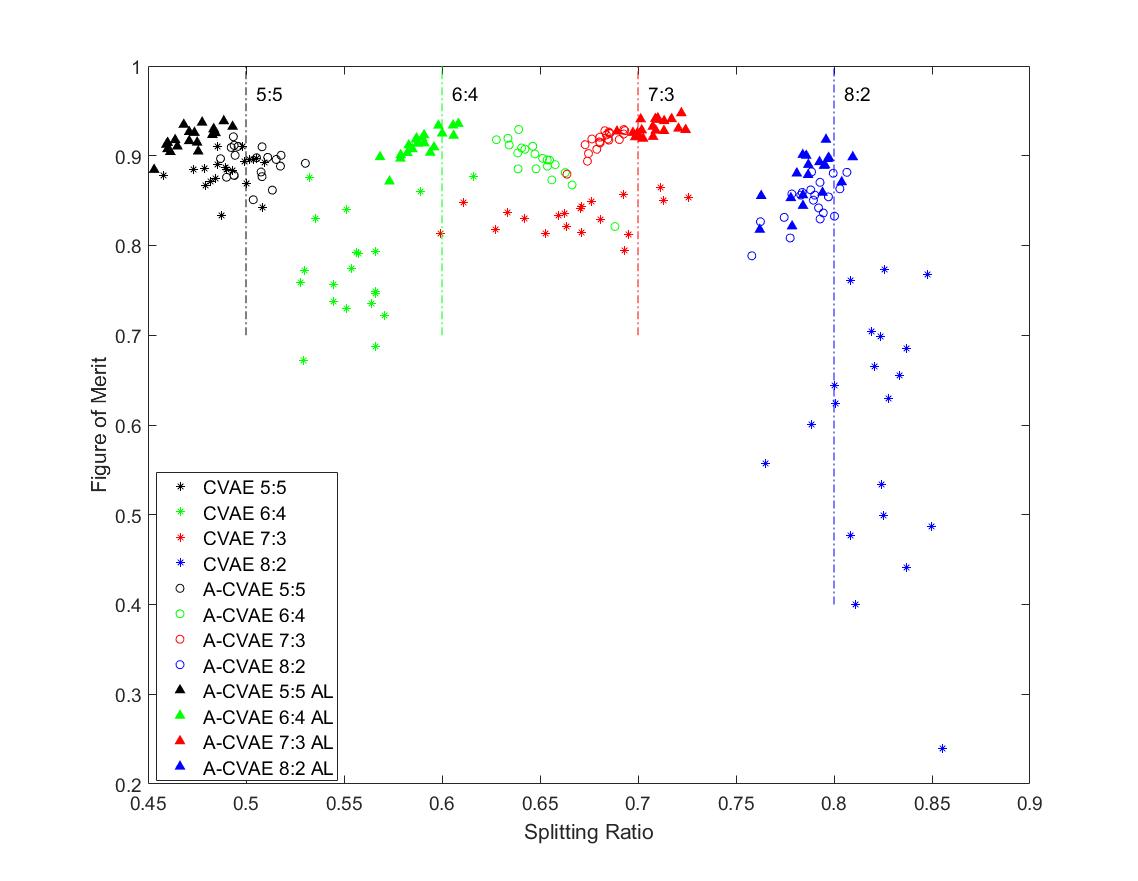}
\caption{FOM  comparison  for  different  CVAE  models:  conventional  CVAE  (star marker),  adversarialCVAE [4] (round marker) and adversarial CVAE with active learning (triangle marker). Four different splitting ratios are used as a target value to test the model performance (marked with dashed lines). The devices generated by the active learning assisted CVAE model can fit the target splitting ratio better with excellent total transmission. The average FOM for the three models are: 0.7705, 0.8877, 0.9009}
\label{fig:FOM-comparison}
\end{figure}

\section*{Performance comparison}

Figure~\ref{fig:FOM-comparison} shows the comparison of the performance among the devices generated by the CVAE model and the devices generated by the CVAE with adversarial censoring. 

The Figure of Merit is calculated for 20 randomly generated devices from the CVAE models and from the best device in the training data. This figure shows that the conventional CVAE model can generally learn the distribution of the data, but it cannot beat the training data in terms of performance. With the help of the adversarial censoring, the generated devices generally have a better performance than the training data. The active learning further improves the performance.

\section*{Computing resource}

Since the network structure is shallower which significantly reduce the training time for the whole system. The batch size that we are using is 128, and optimized iteration number is 1700 and the total training data is a $15000$ binary hole vector pattern set. We are using a Nvidia GTX 1080 GPU and the total training time is around $5$ minutes. 

\section*{Conclusion}

A Conditional Variational Autoencoder (CVAE) with adversarial censoring has been applied to the nanophotonic power splitter application. The Variational autoencoder model takes the binary hole vector as the training sample and can generate patterns with variable hole size. No additional optimization step is needed after the generator generates the pattern. Also, with the help of adversarial censoring, the performance of the generated pattern can be significantly improved (~$5\%$ increase in total transmission). Overall, the device that are generated through our Adversarial Conditional Autoencoder have very good performance (with over $90\%$ in total transmission) across the 550~nm bandwidth. To the authors' knowledge, this is the first demonstration that a CVAE with adversarial censoring was applied to any type of design problem.

\bibliography{main}

\end{document}